\documentclass[sigconf,table,dvipsnames,table,10pt]{acmart}
\acmConference{Position paper}{extended/revised}{based on Eurosys'20DW}
\acmDOI{} 
\acmISBN{} 
\renewcommand\footnotetextcopyrightpermission[1]{}
\setcopyright{none}
\usepackage[utf8]{inputenc}
\usepackage[T1]{fontenc}
\usepackage{natbib}
\usepackage{amsmath,amssymb,amsfonts}
\usepackage{algorithmic}
\usepackage{graphicx}
\usepackage{textcomp}
\usepackage{xcolor}
\usepackage{pifont}
\def\BibTeX{{\rm B\kern-.05em{\sc i\kern-.025em b}\kern-.08em
    T\kern-.1667em\lower.7ex\hbox{E}\kern-.125emX}}

\newboolean{showcomments}
\setboolean{showcomments}{true}
\ifthenelse{\boolean{showcomments}}
{ \newcommand{\mynote}[3]{
   \fbox{\bfseries\sffamily\scriptsize#1}
   {\small$\blacktriangleright$\textsf{\emph{\color{#3}{#2}}}$\blacktriangleleft$}}}
{ \newcommand{\mynote}[3]{}}

\begin{document}

\title{Monitoring Data Distribution and Exploitation\\in a Global-Scale Microservice Artefact Observatory}
\subtitle{Position Paper}

\author{Panagiotis Gkikopoulos}
\affiliation{
  \institution{Zurich Univ. of Applied Sciences}
  \city{Winterthur}
  \country{Switzerland}
  }
\email{pang@zhaw.ch}
\affiliation{
  \institution{University of Neuchâtel}
  \city{Neuchâtel}
  \country{Switzerland}
  }
\email{panagiotis.gkikopoulos@unine.ch}

\author{Josef Spillner}
\affiliation{
  \institution{Zurich Univ. of Applied Sciences}
  \city{Winterthur}
  \country{Switzerland}
  }
\email{josef.spillner@zhaw.ch}

\author{Valerio Schiavoni}
\orcid{0000-0003-1493-6603}
\affiliation{
  \institution{University of Neuchâtel}
  \city{Neuchâtel}
  \country{Switzerland}
}
\email{valerio.schiavoni@unine.ch}


\begin{abstract}
Reusable microservice artefacts are often deployed as black or grey boxes, with little concern for their properties and quality, beyond a syntactical interface description. This leads application developers to chaotic and opportunistic assumptions about how a composite application will behave in the real world. Systematically analyzing and tracking these publicly available artefacts will grant much needed predictability to microservice-based deployments. By establishing a distributed observatory and knowledge base, it is possible to track microservice repositories and analyze the artefacts reliably, and provide insights on their properties and quality to developers and researchers alike. This position paper\footnote{This paper extends the presentation given at EuroDW/Eurosys'20.} argues for a federated research infrastructure with consensus voting among participants to establish and preserve ground truth about the insights.
\end{abstract}

\maketitle

\section{Introduction}

Software is increasingly delivered as a service through clouds and other scalable platforms. Microservices allow for a software application to be developed in a distributed way as well as increasing its resilience and scalability. They typically interact with each other using REST APIs~\cite{Dragoni2017}, message queues or service meshes. Separate teams of developers can work on individual component services of a much larger application \cite{7030212} and some of the resulting software artefacts can be reusable and thus placed in a marketplace for other developers to integrate to other projects.
Beyond existing repositories specific to programming languages such as Maven Central \cite{maven}, RubyGems \cite{ruby} and the Python Package Index \cite{pypi}, microservice-specific repositories, hubs and marketplaces have shown growth over recent months.
These include Docker Hub \cite{docker}, Helm Hub \cite{helm} and the Amazon Serverless Application Repository \cite{awssar} among others.

The widespread adoption of these marketplaces also causes concern as little quantitative data is available to developers, leading to opportunistic design decision with potentially unpredictable results.

We envision the globally operated Microservice Artefact Observatory (MAO~\cite{mao}) as a scientific community effort to monitor and analyze these artefact marketplaces and provide insights through a combination of metadata monitoring, static checks and dynamic testing. For this purpose, we contribute a first working prototype of the corresponding federated research infrastructure for resilient tracking and analysis of marketplaces and artefacts. More importantly, we contribute and outlook of how researchers and developers can benefit from such an approach. A key feature of the infrastructure is a novel method of continuously generating ground truth data resulting from consensus voting over the individual observations and insights. This ground truth is usable by the software engineering and distributed systems communities for further studies.

\textbf{Roadmap.}The reminder of this paper is organized as follows.
First, we survey related work in \S\ref{sec:rw}.
Then, we clarify the problem statement and methodology in \S\ref{sec:prob}.
Some preliminary results are given in \S\ref{sec:res}.
Finally, we describe our future work in \S\ref{sec:future}.

\section{Related Work}\label{sec:rw}
Various works have delved into the monitoring and testing of quality aspects of microservice architectures.
Some approaches focus on the metadata and logs generated by the microservices~\cite{8103476}, rather than benchmarking specific architectures~\cite{DBLP:journals/corr/abs-1807-10792,7133548} or testing dependencies and service interactions \cite{8377834}.
Researchers looked also at runtime monitoring of microservice-based applications \cite{8377902}.
More recently, there were attempts~\cite{Gan:2019:OBS:3297858.3304013} for benchmarking architectural models based on microservices to study the implications of the design pattern on real world applications.
\textsc{Sieve}~\cite{DBLP:journals/corr/abs-1709-06686} extracts usable metrics for developers by monitoring these software artefacts.
A similar approach was proposed in~\cite{7958458} for corporate managers.
Finally, there are efforts to define a standard set of requirements for the orchestration of microservices \cite{10.1007/978-3-319-74781-1_16} as well as surveys to detect trends in microservice development \cite{DBLP:journals/corr/abs-1808-04836}.

We observe the lack of an approach to systematic monitoring of microservice-related marketplaces. Current tools and platforms are aimed at analyzing and monitoring specific case studies and architectures to a significant depth, but the publicly available reusable artefacts are, for the most part, not scrutinized for their quality properties nor tracked with respect to potential improvements or regressions.
We argue that such an approach is necessary for providing preemptive quality metrics to developers that aim to reuse publicly available artefacts, as well as researchers studying the evolution of the field and emerging trends and issues.

Prior work include a data crawler of Helm charts on the Kube Apps Hub~\cite{DBLP:journals/corr/abs-1901-00644}, on the Serverless Application Repository \cite{DBLP:journals/corr/abs-1905-04800} for QA and preservation purposes, as well as \emph{DApps} across multiple marketplaces~\cite{dapps}.
The current output of this effort includes static analysis software tools, as well as experimental datasets~\cite{sardata, helmdata, dappsdata} in addition to the research papers and preprints.
We intend to integrate and upstream the efforts and vision presented in this paper into these open initiatives with the goal that future metrics collections no longer rely on standalone, brittle and centralised tools.

A clear gap we have identified in the state of the art is a large scale monitoring system with the capability to track the evolution of these marketplaces and the artefacts within, to provide both broad and deep insight on the state of the ecosystem.

\section{Overview}\label{sec:prob}
\subsection{Problem Statement}
\begin{figure}[!t]
    \centering
    \includegraphics[width=\columnwidth]{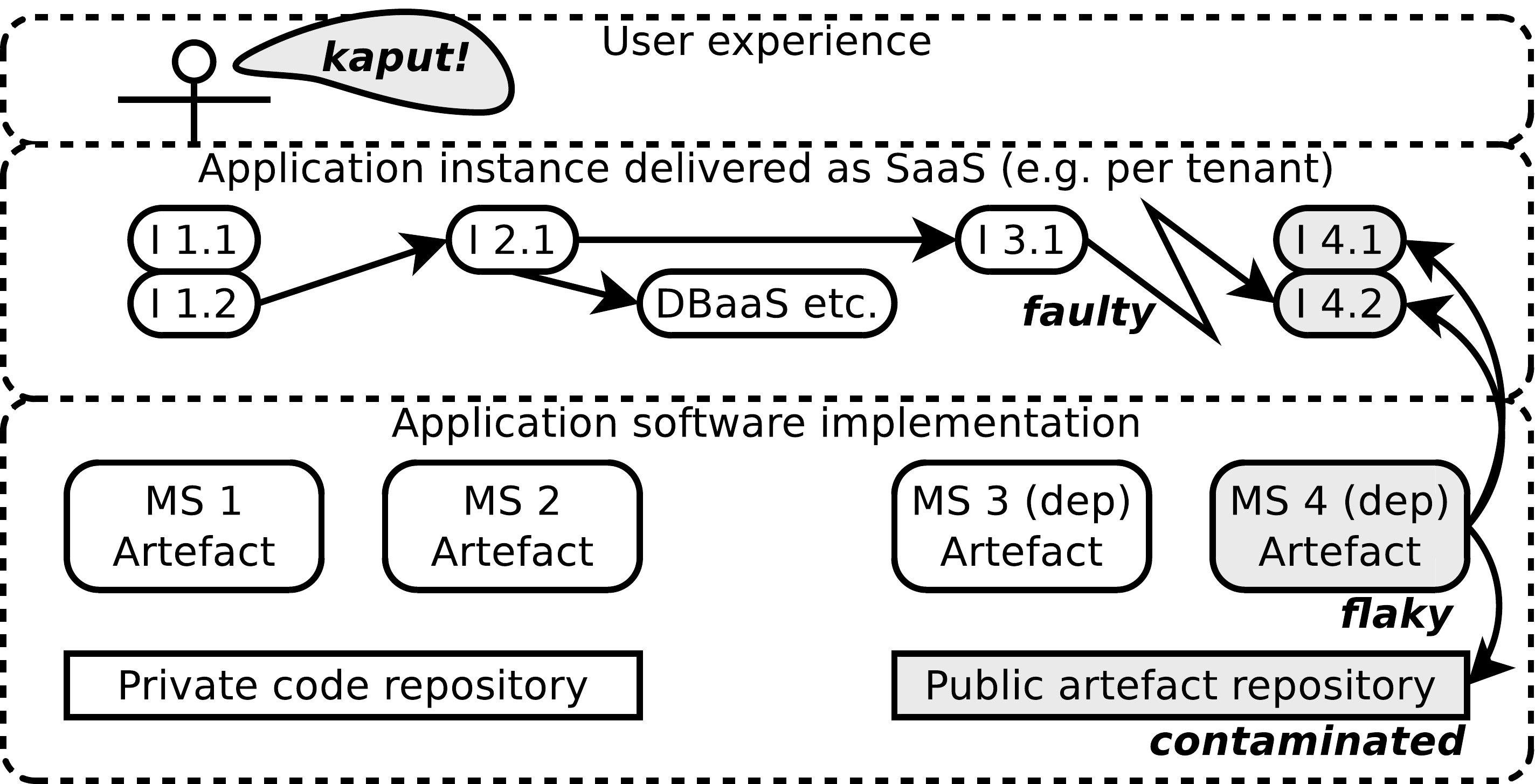}
    \caption{Propagation of weaknesses in microservice artefacts to the application-level quality of experience}
    \label{fig:composition}
\end{figure}

Treating reusable microservice artefacts as black or grey boxes when designing a complex application can lead to wrong assumptions about their properties and thus unpredictable behavior of the application itself. Small mistakes in a microservice implementation or configuration, even in dependency third-party artefacts, might endanger the entire application as shown in Fig. \ref{fig:composition}. The quality of compositions is after all limited by their weakest point. Providing information in the form of metadata analysis, code quality, performance benchmarking and security evaluation for these artefacts can increase the predictability of their behaviour in production by allowing the developers to make informed design decisions and be aware of potential issues ahead of time. This would also help researchers in the field have access to insights on development trends and emerging anti-patterns in microservice architectures.

Additionally, current methods used to monitor software repositories lack infrastructure to enable large scale collaborations and robust historical tracking. Yet, we argue that such infrastructure is a requirement for monitoring projects of this scale, especially since an international consortium (albeit an informal one in this case) is involved. We identify that the capability of reliable continuous tracking within a decentralized system as well as an algorithm or set of algorithms to establish ground truth when multiple data snapshots are submited by peers within the collaboration as the key features of the proposed infrastructure.

Our approach to this problem is focused around the following research questions:

\begin{itemize}
    \item \textbf{RQ 1:} How can a distributed, federated system enable more efficient and resilient monitoring and analysis of microservice artefacts at the scale of a marketplace?
    \item \textbf{RQ 2:} How can cluster consensus be utilised within the federation to establish ground truth about artefact quality metrics?
\end{itemize}

\subsection{Method}
We envision the work to be pursued along two main dimensions: \emph{(1)} the establishment and engineering of the proposed observatory infrastructure, and \emph{(2)} the exploitation of said infrastructure within our primary use case of collecting artefact metrics.

The engineering component involves providing the functionality needed to assist and automate all aspects of the data management pipeline, from scheduling data acquisition tools to comparing snapshots of the data from different nodes to reach a ground truth measurement. Additionally, a main requirement is to provide resilience and reliability, to protect from hardware outages and corrupt data files.

\begin{figure*}[ht]
    \centering
    \includegraphics[scale=0.7]{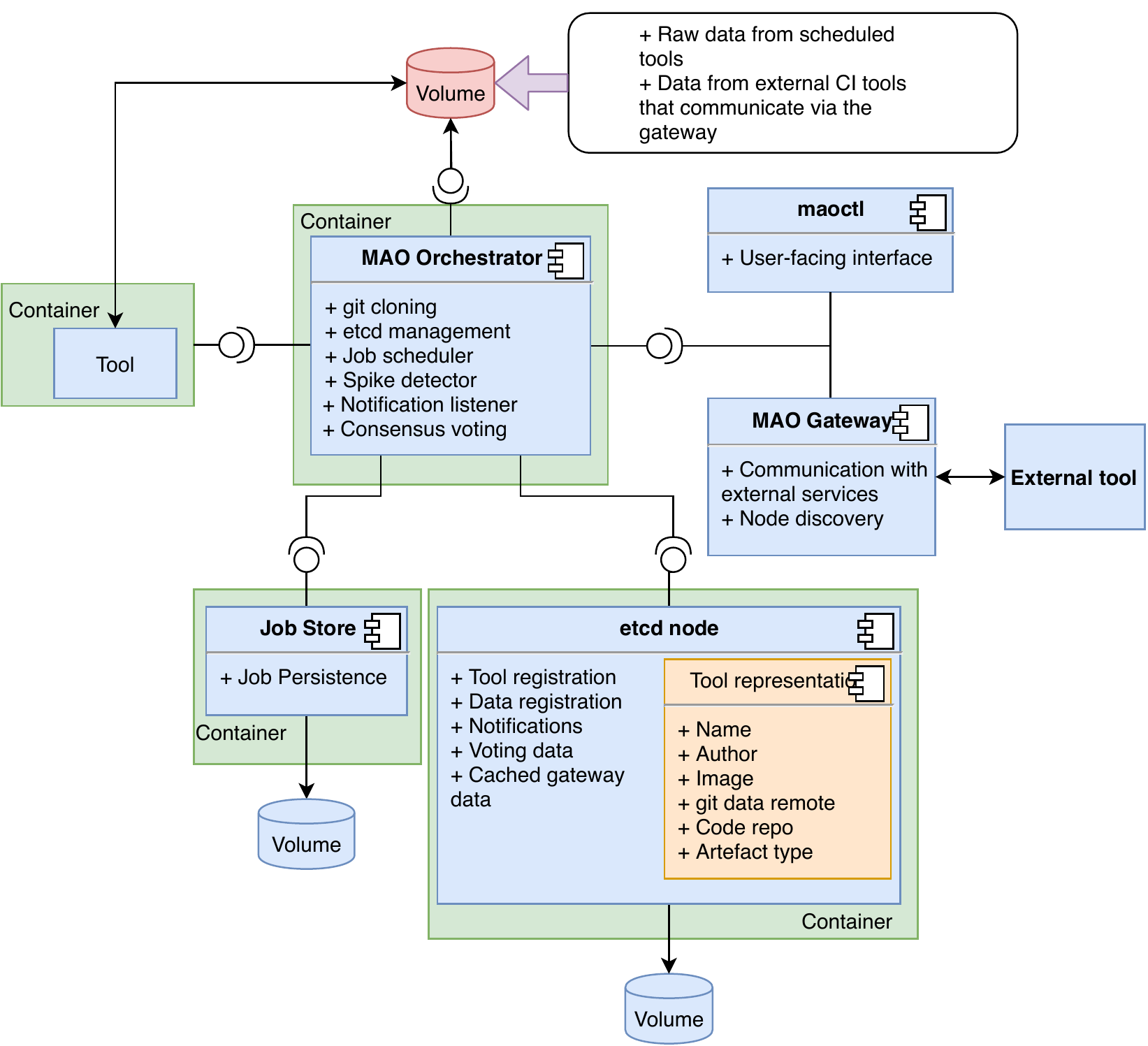}
    \caption{Current observatory architecture}
    \label{fig:architecture}
\end{figure*}

Fig. \ref{fig:architecture} shows the current architecture for the orchestration system. It primarily consists of an orchestration/scheduling service that manages the data acquisition tools. Nodes access an etcd \cite{etcd} cluster to share registry information such as available tools to deploy, dataset repositories and notifications. An additional gateway service can also be deployed to connect external tools if needed.

\begin{figure*}[ht]
    \centering
    \includegraphics[scale=0.7]{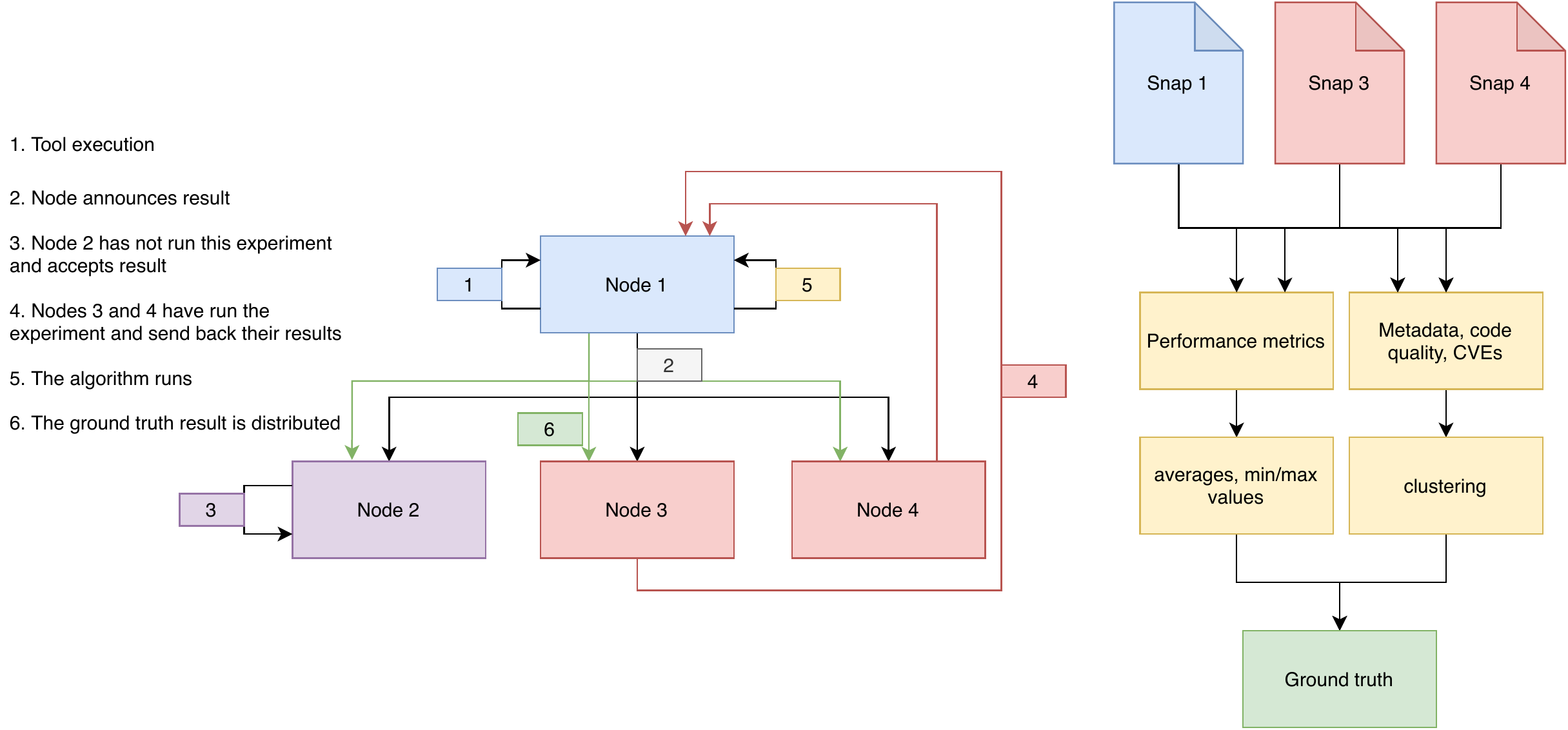}
    \caption{Consensus voting concept diagram}
    \label{fig:voting}
\end{figure*}

Apart from data acquisition automation the orchestration system aims to assist in collaborations by partially automating the replication of experiments and the verification of results. Fig. \ref{fig:voting} shows the current concept algorithm that will be implemented in the next stages.
A node that runs an experiment can announce its results to other nodes (Fig.~\ref{fig:voting}-\ding{203}).
If the nodes have not executed this data acquisition tool (Fig.~\ref{fig:voting}-\ding{204}), they will simply accept this measurement.
However, if they already have a measurement (Fig.~\ref{fig:voting}-\ding{205}), they will respond to the announcement, indicating a comparable data snapshot in the \texttt{etcd} registry.
The first node will then retrieve these snapshots from their respective repositories (\emph{e.g.}, git) and run the verification algorithm (Fig.~\ref{fig:voting}-\ding{206}).

The verification makes a distinction between performance metrics and quality or security metrics.
In the case of benchmarking or other performance metrics (as labeled in the dataset itself) the verification will produce average, minimum and maximum values as the ground truth measurement. For metrics such as metadata or vulnerability characteristics, where average values would be meaningless, the ground truth measurement will be arrived at via clustering.

Once the verification is complete and the ground truth data snapshot is established, it will once again be announced to all nodes (see Fig.~\ref{fig:voting}-\ding{207}), this time marked as verified so nodes can accept it unless a more recent measurement has surfaced.

The second dimension will focus on the exploitation of the new architecture within the MAO use case in order to both further the goals of MAO of reliable tracking of the evolution of microservice artefact repositories, as well as evaluate the improvement this federated system brings to the current state-of-the-art research practice. As the observatory grows, with more tools and metrics and more collaborators, so will out view of the system's effectiveness increase, allowing us to better gauge how it scales when applied to a large real-world scenario. For this purpose, we have initiated a free collaborative network of researchers and pilot software engineers\footnote{MAO collaboration: \url{https://mao-mao-research.github.io/}}.

\section{Preliminary Results}\label{sec:res}
Early contributions in this work have been divided between improving the infrastructure of the experiments, and developing additional tooling for monitoring and testing artefacts.

We developed a distributed monitoring architecture using a geo-distributed cluster using \texttt{etcd} for metadata exchange among peers and a Docker-based scheduler-orchestrator application for members to run the monitoring tools~\cite{orchestrator}.
The system is currently operated in our own servers in Switzerland, with a pilot deployment in Argentina and additional deployments being discussed with MAO-member researchers.

\begin{figure*}
    \centering
    \includegraphics[scale=0.39]{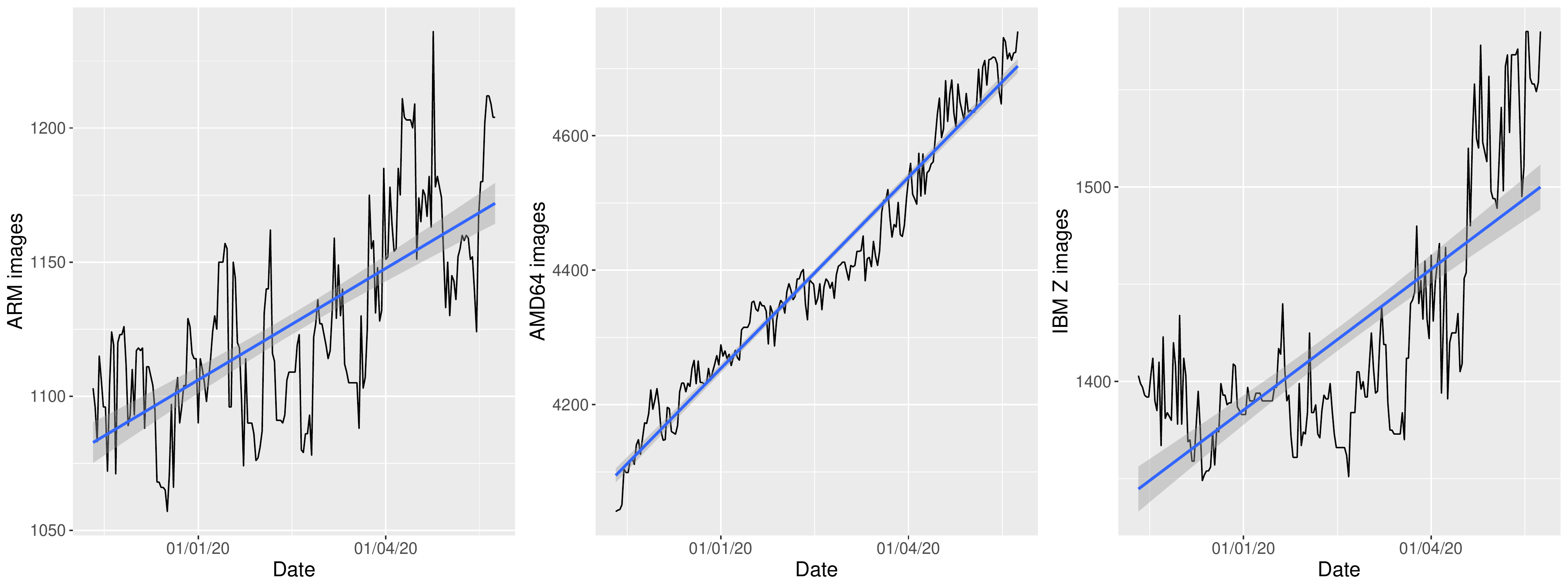}
    \caption{Dockerhub monitoring for Nov 11, 2019 to May 24, 2020}
    \label{fig:my_label2}
\end{figure*}

The first monitoring tool deployed on the system is an automatic crawler of Dockerhub's public API.
It collects image metadata and provides basic insights on the development trends within the ecosystem~\cite{collector}.
The first version focused on OS support and CPU architecture for each image, to better understand the extent of Docker's support for heterogeneous ecosystems.
Figure~\ref{fig:my_label2} shows the current set of data produced by the crawler. We tracked the evolution of support for different CPU architectures over time. The graph shows the number of images for ARM, x86-64 and IBM Z architectures for each date of the tracking period, along with a trend line to highlight increase/decrease over time.

A concurrent experiment focuses on extending the existing study of the AWS Serverless Application Repository with a benchmarking tool, to test the collected artefacts.
To that end, an emulation based on LocalStack~\cite{localstack} and \texttt{sam local}~\cite{samlocal} is currently under development.
The aim is to emulate the behaviour of AWS's serverless offering to provide more accurate metrics than current unit-test tools.
We also quantitatively assess Dockerfiles with multiple linters on the source level.

\section{Conclusion and Future Work}\label{sec:future}
We proposed an architecture to analyze data from monitored artefact marketplaces.
This architecture is currently being extended to support dynamic execution of the gathered artefacts, and to allow on-demand comparison of data between nodes, to establish ground truth data via cluster consensus.

The analysis will be extended towards a more diverse set of artefacts types, such as, such as Kubertetes Helm Charts, Docker images and Compose files, Kubernetes Operators and others, furthering the goals of the MAO project while simultaneously allowing us to evaluate the observatory architecture at scale.

Additionally, we aim to integrate the knowledge base we build over time into data-driven QA tooling, that can be used within CI/CD pipelines, thus providing real-time quality analysis and feedback to developers.

\section*{Acknowledgments}

We thank Paolo Costa, John Wilkes and reviewers of the original paper for feedback and suggestions leading to this extended presentation of our position and approach.
\\

\end{document}